\documentclass[aps,showpacs,twocolumn,prl,superscriptaddress]{revtex4-1}
\usepackage{graphicx}
\usepackage{amsfonts}
\usepackage{color}

\begin{document}

\title{Stochastic Loewner Evolution Relates Anomalous Diffusion and Anisotropic
       Percolation}

\author{Heitor F. Credidio}
\email{credidio@fisica.ufc.br}
\affiliation{Departamento de F\'isica, Universidade Federal do Cear\'a,
             Campus Do Pici, 60455--760 Fortaleza, Brazil}

\author{Andr\'e A. Moreira}
\email{auto@fisica.ufc.br}
\affiliation{Departamento de F\'isica, Universidade Federal do Cear\'a,
             Campus Do Pici, 60455--760 Fortaleza, Brazil}

\author{Hans J. Herrmann}
\email{hans@fisica.ufc.br}
\affiliation{Departamento de F\'isica, Universidade Federal do Cear\'a,
             Campus Do Pici, 60455--760 Fortaleza, Brazil}
\affiliation{Computational Physics IfB, ETH Zurich, Stefano-Franscini-Platz 3,
             CH-8093 Zurich, Switzerland}

\author{Jos\'e S. Andrade Jr.}
\email{soares@fisica.ufc.br}
\affiliation{Departamento de F\'isica, Universidade Federal do Cear\'a,
             Campus Do Pici, 60455--760 Fortaleza, Brazil}
\affiliation{Computational Physics IfB, ETH Zurich, Stefano-Franscini-Platz 3,
             CH-8093 Zurich, Switzerland}

\date{\today}

\begin{abstract}
    We disclose the origin of anisotropic percolation perimeters in terms of
    the Stochastic Loewner Evolution (SLE) process. Precisely, our results from
    extensive numerical simulations indicate that the perimeters of
    multi-layered and directed percolation clusters at criticality are the
    scaling limits of the Loewner evolution of an anomalous Brownian motion,
    being subdiffusive and superdiffusive, respectively. The connection between
    anomalous diffusion and fractal anisotropy is further tested by using
    long-range power-law correlated time series (fractional Brownian motion) as
    driving functions in the evolution process. The fact that the resulting
    traces are distinctively anisotropic corroborates our hypothesis. Under the
    conceptual framework of SLE, our study therefore reveals new perspectives
    for mathematical and physical interpretations of non-Markovian processes in
    terms of anisotropic paths at criticality and vice-versa.
\end{abstract}

\pacs{89.75.Da, 64.60.al, 05.40.Jc}
\maketitle

\newcommand{\hpl}{\ensuremath{\mathbb{H}}}
\newcommand{\cpl}{\ensuremath{\mathbb{C}}}
\newcommand{\red}[1]{\textit{\textcolor{red}{#1}}}
\newcommand{\rr}{\textit{\red{[refs]}}}
\newcommand{\re}{\ensuremath{\mbox{Re}}}
\newcommand{\im}{\ensuremath{\mbox{Im}}}
\newcommand{\p}[1]{\ensuremath{\left(#1\right)}}


The Stochastic Loewner Evolution (SLE)~\cite{Schramm2000} has revolutionized
our understanding of two dimensional loopless paths, as recognized among others
by several Fields medals~\cite{Mackenzie2006, Kesten2010}. It provides a
mapping between these paths and a real valued function, called ``driving
function'', that is a random walk if the path is a conformally invariant
fractal. This establishes a relation between the fractal dimension of the path
and the diffusion constant of the random walk. Although several generalizations
have been proposed~\cite{Rushkin2006, Oikonomou2008, Zhan2004, Najafi2012}, due
to its nature, SLE has been restricted to isotropic models~\cite{Smirnov2001,
Lawler2002, Kager2004, Cardy2005, Lawler2011, Daryaei2012, Pose2014}.
However, anisotropic paths, namely, paths with a preferential direction,
appear quite commonly in Physics. By numerically determining the driving
function of anisotropic paths, we discover that they are consistently
mapped onto correlated random walks, meaning that the Markovian property of
the driving function is violated. More precisely, we show that resulting
anomalous diffusion is characterized by an exponent that is related to the
degree of anisotropy. This behavior can be subdiffusive, as it is the case
for the hull of directed percolation or superdiffusive, as found for
multi-layered percolation.

\begin{figure}[b]
    \includegraphics[scale=0.2]{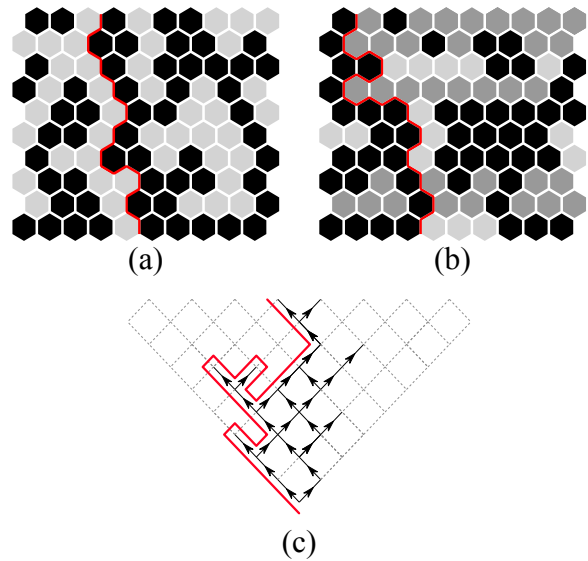}
    \caption{\label{fig1}
        Percolation models used to generate the SLE curves. (a) Regular
        percolation, where each site is occupied with the same probability
        $p$~\cite{Stauffer1994}. (b) Multi-layered percolation, where some rows
        are occupied with probability $p+\Delta$ (dark gray rows) and others
        with $p-\Delta$ (light gray rows)~\cite{Dayan1991}. (c) Directed
        percolation is a spreading process which starts at the bottom of
        the tilted lattice and can only advance upwards with probability
        $p$~\cite{Hinrichsen2000}.
    }
\end{figure}

\begin{figure*}
    \includegraphics[scale=0.2]{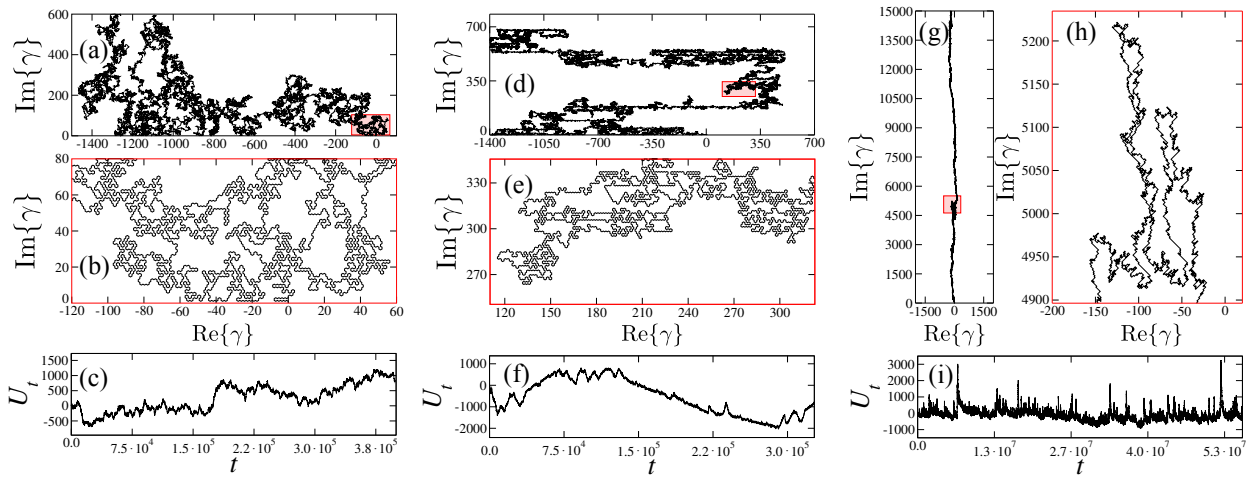}
    \caption{\label{fig2}
        Examples of cluster perimeters at the critical point of (a)
        isotropic percolation on a triangular lattice ($p_c=0.5$), (d)
        multi-layered percolation also on a triangular lattice ($\Delta=0.2$
        and $p_c=0.5$), and (g) directed percolation on a square lattice
        ($p_c\approx0.644$). A detail of each curve in (a), (d) and (g) (pink square) can be
        seen in (b), (e) and (h), respectively. The driving functions obtained
        by applying the zipper algorithm~\cite{Kennedy2008} to the curves (a),
        (d) and (g) are shown in (c),  (f) and (i), respectively.
    }
\end{figure*}

The \textit{chordal} variety of SLE (the one we will focus on this work)
deals with curves $\gamma_t$ that start at the origin and grow towards infinity
while restricting themselves to the complex upper half-plane $\hpl$. The curve
$\gamma_t$ is connected to a real-valued driving function $U_t$
through the relation $\gamma_t = g_t^{-1}(U_t)$ where $g_t(z)$ is
the solution of Loewner's equation~\cite{Loewner1923, Lawler2008},
\begin{equation}
    \label{eq:loew}
    \partial_t g_t\p{z} = \frac{2}{g_t(z) - U_t},\;\;\;\;g_0(z)=z.
\end{equation}

In his seminal work~\cite{Schramm2000}, Schramm showed that if the measure over
$\gamma_t$ displays conformal invariance and domain Markov property, then the
only possibility is that $U_t$ be a Brownian motion with a single free
parameter $\kappa$, the diffusion coefficient. This is often written as $U_t =
\sqrt{\kappa}B_t$ where $B_t$ is a standard Brownian motion (with diffusion
coefficient equal to unity). The value of $\kappa$ is related to the geometric
properties of $\gamma_t$, including its fractal dimension, which is determined
by the relation $d_f=\min\p{1+\frac{\kappa}{8},2}$~\cite{Beffara2008}. A few
lattice models have been shown to converge to SLE in the continuum limit
\cite{Schramm2000, Lawler2011, Smirnov2007}, and many more are conjectured to
do so~\cite{Lawler2002, Bernard2007, Bogomolny2007, Gamsa2007, Daryaei2012,
Pose2014}. Of particular interest is the proof that the perimeter of a
percolation cluster on a triangular lattice follows SLE with $\kappa = 6$ as a
scaling limit, allowing for a formal computation of its critical
exponents~\cite{Smirnov2001}.

In this work we explore the possibility of using Loewner evolutions to study
anisotropic fractal systems, i.e., systems with different critical exponents in
each direction. These systems are not scale invariant, therefore not
conformally invariant either. We are particularly interested in two variants of
the percolation model that show anisotropic behavior, namely, multi-layered
percolation and directed percolation (see Fig.~\ref{fig1}). Precisely,  we
generate the border of percolating clusters, numerically compute their
corresponding driving function, and then analyze the diffusive
properties of these numerical sequences. In the general case, we expect that the
mean square displacement of $U_t$ behaves like,
\begin{equation}
    \label{eq:diff}
    \left\langle U_t^2 \right\rangle \rightarrow b t^\alpha
\end{equation} 
as $t\rightarrow\infty$. In the case of traditional SLE, $\alpha=1$
and $b=\kappa$.  The driving functions of anisotropic percolation
models, we found, display very distinctive anomalous diffusive
behavior ($\alpha\neq1$). Finally, we show that our approach is also valid in
the opposite direction, namely, the SLE consistently leads anomalously
diffusive driving functions to traces that display clear anisotropic scaling.

In order to evaluate the SLE driving function of the cluster perimeters we used the
zipper algorithm with a vertical slit discretization~\cite{Bauer2003, Kennedy2008}.
In this method, given a lattice curve $\{0, \gamma_1, \gamma_2, \ldots,
\gamma_N\}$, its driving function can be recovered by applying the relations,
\begin{equation}
    \label{eq:zip1}
    t_k = \frac{1}{4}\sum_{j=1}^{k}\im\{\omega_j\} ^ 2
    \;\;\;\;\;
    U_{t_k} = \sum_{j=1}^k\re\{\omega_j\},
\end{equation}
where the $\omega_k$'s are determined recursively by
\begin{equation}
    \label{eq:zip2}
    \omega_k = f_{k-1}\circ f_{k-2}\circ\ldots\circ f_{1}(\gamma_k)
    \;\;\;\;\;
    \omega_1=\gamma_1,
\end{equation}
and $f_k(z)$
\begin{equation}
    \label{eq:zip3}
    f_k(z) = i \sqrt{-\im\{\omega_k\} ^ 2 - {(z - \re\{\omega_k\} )} ^ 2}.
\end{equation}
This algorithm, however, does not guarantee that the discretized times
$t_k$ are equally distributed, even for curves of same length and step size. To
obtain an ensemble of curves defined for the same time sequence, we linearly
interpolate the obtained driving function at equally spaced points in
logarithmic time in the interval  $[1, \log t_f]$, for some suitable $t_f$.

As already mentioned, we also performed the opposite operation of computing
the SLE trace from a given driving function The process is simply the inversion
of the algorithm previously described. Given a discretized driving function
$U_t = \{0, U_{t_1},
\ldots, U_{t_N}\}$, the trace can be obtained by repeatedly applying the
functions
\begin{equation}
    \label{eq:unzip1}
    \gamma_i = g_0 \circ g_1 \circ \ldots \circ g_i(0),
\end{equation}
where the mappings are also chosen to represent a vertical slit discretization,
\begin{equation}
    \label{eq:unzip2}
    g_i(z) = i\sqrt{4 {(t_i - t_{i-1})}^2 - z^2} + (U_{t_i} - U_{t_{i-1}}).
\end{equation}

Instead of using an approximate algorithm~\cite{Kennedy2007}, we chose to use
the one described above, as they are exact (for a given discretization). Their
complexity scales as $O(N^2)$ which can get quite time consuming, specially for
large values of $\kappa$, requiring a large number of points to get accurate
results. We resorted to GPU parallelization (where each $\gamma_k$ is computed
by a single thread) to achieve satisfactory accuracy.

\begin{figure}
    \includegraphics[scale=0.27]{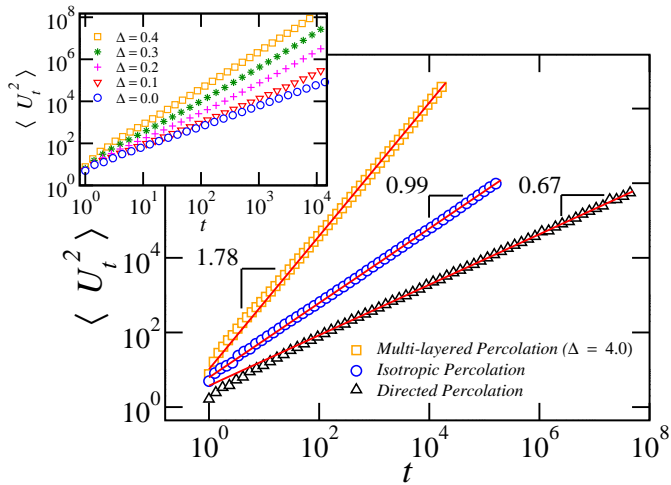}
    \caption{\label{fig3}
        Mean squared displacement of the driving functions for the three
        percolation models studied. The curves are the results of the numerical
        procedure described in the text applied to $10^4$ realizations of each
        type of percolation model. The $95\%$ confidence intervals were
        bootstrapped over $400$ resamplings~\cite{Felsenstein1985}, but, being
        smaller than the symbols, are not shown. As expected, in the case of
        isotropic percolation, the displacement scales linearly with time,
        while it shows instead a distinctive subdiffusive behavior for directed
        percolation, with an exponent $\alpha\approx 0.67$. In the case of
        multi-layered percolation, a clear superdiffusive behavior, with an
        exponent $\alpha\approx 1.78$, can be observed for $\Delta=0.4$. The
        inset shows how this anomalous diffusion regime is gradually achieved
        as we increase the degree of anisotropy $\Delta$.
    }
\end{figure}

We start by testing our approach on standard isotropic percolation,  which has
been extensively studied as a simple but rather rich and illustrative model for
criticality~\cite{Stauffer1994}. It is basically a lattice model with binary
disorder, where each site (or bond) is occupied with probability $p$. For a
given critical probability $p_c$, the presence of a giant spanning cluster is
detected. In the thermodynamic limit, if $p<p_c$, where $p_c$ is the
percolation threshold, the system never percolates, otherwise it always does.
In particular, $p_c=1/2$ for site percolation on the triangular
lattice~\cite{Stauffer1994}. It has been mathematically proven that the
perimeter of the giant cluster at the critical point~\cite{Camia2006} follows
SLE with $\kappa= 6$~\cite{Smirnov2001}. We perform simulations with $10^4$
realizations of percolation perimeters of length $10^5$ lattice units generated
using the algorithm described in Ref.~\cite{Ziff1984} on the triangular
lattice. Fixed boundary conditions are adopted, in which every site on the left
side of the bottom row is always unoccupied and the ones on the right side are
always occupied.  In Fig.~\ref{fig2}a we show a typical realization of an
isotropic percolation perimeter and the corresponding driving function, as
computed using the algorithm Eqs.~\ref{eq:zip1}-\ref{eq:zip3}. Finally, from
the driving functions, we calculate their mean squared displacement $\langle
U_{t}^{2} \rangle$ as a function of time, and find that $\kappa=6.27\pm0.30$
and $\alpha=0.996\pm0.005$, as shown in Fig.~\ref{fig3}a.

Unlike regular percolation, where each site or bond is occupied with
probability $p$, in multi-layered percolation this is done with probability
$p\pm\Delta$, where $\Delta\in[0,\frac{1}{2}]$, and the signs plus or minus are
chosen randomly with equal probabilities for each row of the
lattice~\cite{Dayan1991, Parteli2010}. Here, the parameter $\Delta$ represents
the degree of anisotropy of the system, with $\Delta=0$ being equivalent to
isotropic (regular) percolation.

We generate an ensemble of $10^4$ multi-layered percolation perimeters  of
length $10^5$ lattice units on a triangular lattice at the critical point for
different values of $\Delta$. For every value of $\Delta$ we found $p_c=0.5$
using the cluster perimeter method~\cite{Ziff1986}. As for the standard
percolation case, after calculating the driving functions, an example of which
is shown in Fig.~\ref{fig2}b, we compute the corresponding mean
square-displacement to  find that it exhibits characteristic superdiffusive
behavior for every value of $\Delta>0$. As can be observed in the inset of
Fig.~3, however, a long transient behavior is present for small values of
$\Delta$ before a distinctive power-law behavior is established. For
$\Delta=0.4$, after a short transient, the least-squares fit to the data gives
a power-law, $\langle U_{t}^{2} \rangle=bt^{\alpha}$, with $b=10.38\pm0.68$ and
$\alpha=1.78\pm0.01$, which extends over more than three orders of magnitude. 

Next we investigate the diffusive behavior of driving functions generated from
directed percolation perimeters.  As defined, directed percolation is a
spreading  process where a cluster can only grow along preselected directions
in a lattice, and each site is occupied with probability
$p$~\cite{Hinrichsen2000}. Shown in Fig.~\ref{fig1}c is a typical realization
of a directed percolation perimeter generated on a tilted square lattice at
the critical point, $p_c=0.644700185(5)$~\cite{Jensen1999}. Using this
simulation setup, the perimeters of the spanning clusters are obtained here
using a simple walker algorithm, as illustrated by the red curve in the
example shown in Fig.~1c. From the ensemble of the generated driving functions,
once more the resulting mean square displacement displays a characteristic
anomalous behavior. Precisely, the least-squares fit to the data in the scaling
region yields subdiffusive diffusion, as shown in Fig.~3, with a pre-factor
$b=3.74\pm0.07$ and an exponent $\alpha=0.676\pm0.001$.

The results obtained by the previous analysis suggest that the presence of
long-range correlations in the driving function should lead, through the
Loewner evolution process, to anisotropic fractal traces, and vice-versa. In
order to test this hypothesis, we analyze the behavior of traces driven by
stochastic processes exhibiting anomalous diffusion. We choose to use
fractional Brownian time series generated according to a given Hurst exponent
$H$, which is related to the diffusion exponent by
$\alpha=2H$~\cite{Mandelbrot1968}.

We generated the drive $U_t$ as a fractional Brownian Motion with Hurst
exponent $H$ and diffusion constant $b$ in $N$ time steps $t_i$ uniformly
spaced in the interval $[0, t_f]$. In order to simulate fractional Brownian
motions with reasonable control over the diffusive constant $b$, the
Davies-Harte algorithm was used~\cite{Davies1987}. The $\gamma_{t_i}$ were
computed from $U_{t_i}$ using Eq.~\ref{eq:unzip1}. We then interpolated the
trace $\gamma(\ell)$ (the same $\gamma_{t_i}$ as before, but parametrized by
its length instead of the Loewner time) in $M$ equally spaced points
$\ell_i\in[0,\ell_{max}]$. This interpolation step was necessary because the
zipper algorithm generates discretized traces with highly non-uniform step
sizes $|\gamma_i-\gamma_{i-1}|$. Although this does not diminish the intrinsic
error of the algorithm, it makes the analysis easier to perform. In order to
study whether the scaling is isotropic or anisotropic, the root mean squared
estimation of the displacement of the trace was computed in each direction,
that is,
\begin{equation}
    \label{eq:correl}
    F_X(i\Delta\ell) = \sqrt{\frac{1}{M-i}
    \sum_{j=0}^{M-i}{[X(\ell_{j+i}) - X(\ell_j)]}^2 },
\end{equation}
where $X(\ell) = \re\{\gamma(\ell)\}$.
Analogously, $F_Y(i\Delta\ell)$ is defined taking instead
$Y(\ell)=\im\{\gamma(\ell)\}$.

\begin{figure*}
    \centering
    \includegraphics[width=1.0\linewidth]{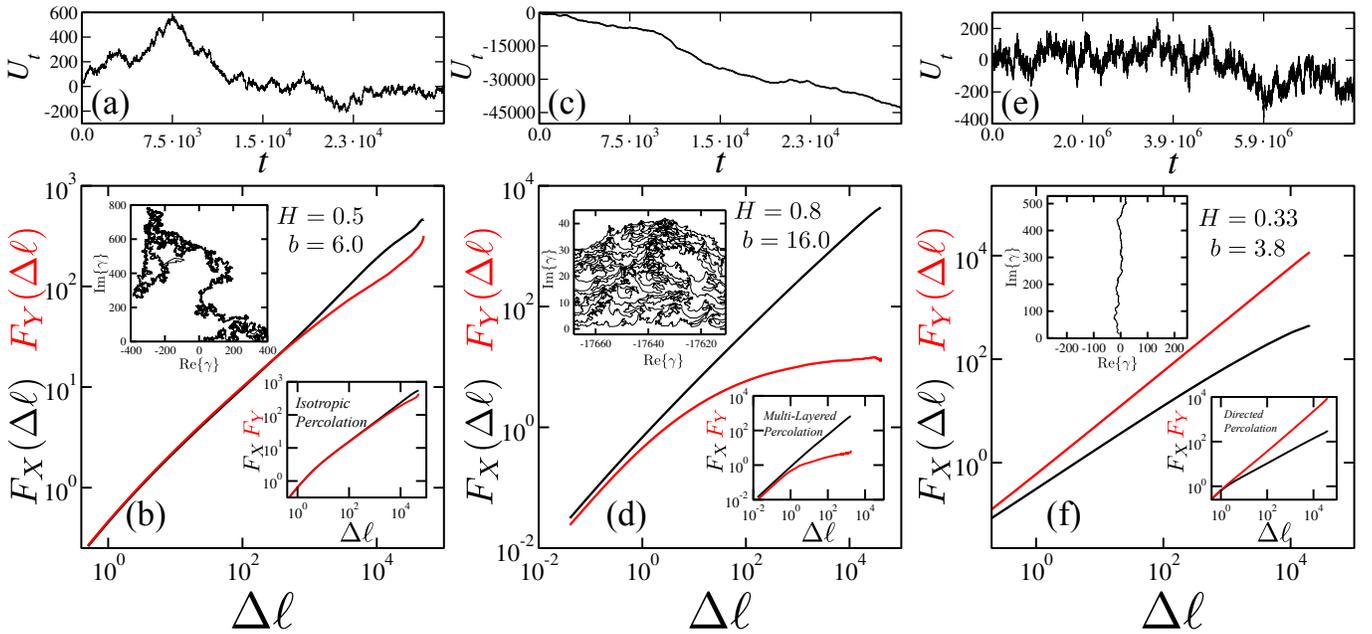}
    \caption{\label{fig4}
        Root mean squared estimations of the displacements in the $X$ and
        $Y$-directions of SLE traces driven by long-range power-law correlated
        time series (fractional Brownian motion). In (a), (c) and (e) we show
        typical realizations of uncorrelated, correlated and anti-correlated
        driving functions, respectively. The simulation parameters ($H$, $b$
        and $t_f$) were chosen based on the results shown in Fig.~\ref{fig3}
        (see Table~\ref{tab1} for the numerical values). Good agreement is
        observed between the uncorrelated result (b) and isotropic percolation
        (inset on the bottom), as it is expected. The correlated trails (d) are
        also compatible with multi-layered percolation (inset on the bottom).
        In the anti-correlated case (f), the same kind of anisotropy present in
        the directed percolation is observed (inset on the bottom). These
        results support our hypothesis that long-term correlations in the
        driving functions, i.e., the presence of anomalous diffusion, are
        responsible for the anisotropic behavior of the traces. The insets on
        the top of (b), (d) and (f) show examples of the traces generated from
        the simulations with the corresponding driving function shown in (a),
        (c) and (e), respectively.
    }
\end{figure*}
Our numerical scheme was applied to three sets of times series, each with
100 realizations generated to reproduce the corresponding properties (in terms
of $H$, $b$ and $t_f$) of the driving functions originated  from the isotropic
and anisotropic percolation traces previously investigated. More precisely, the
first set of time series corresponds to uncorrelated Brownian  motion, the
second to correlated or persistent, and the third is anti-correlated  or
anti-persistent~\cite{Peng1993}. The remaining parameters ($N$, $M$
and $\ell_{max}$) are chosen to ensure the accuracy of our results. The precise
values of all parameters adopted in the simulations are reported in Table I.
Figure~\ref{fig4} shows that the traces evolved from these time series have
similar behavior to their corresponding percolation models, in the sense that a
clear anisotropy can be observed in the correlated and anti-correlated
simulations,  while the uncorrelated one displays isotropic behavior, as
expected.

In summary, our numerical analysis offers compelling evidence that a variation
of the Stochastic Loewner Evolution, obtained by taking as driving function a
stochastic process with anomalous diffusion, may be the scaling limit of
anisotropic critical models. In particular we looked at two anisotropic
variants of percolation: directed percolation and multi-layered percolation.
The former was found to be associated with subdiffusive driving functions,
while the latter are superdiffusive. We also tested the inverse
relation, finding that driving functions with anomalous diffusion do indeed
generate traces with anisotropic features.  This possibility opens new
questions, like how the critical exponents of SLE traces depend on the addition
of long-term correlations to the driving function. Moreover it would be
interesting to know if one can obtain exponents of actual physical models with
such a generalized theory. We expect that the further developments of this new
variant of SLE may provide some insight on the critical behavior of anisotropic
systems, the same way the original SLE was to isotropic systems.

\begin{table}
    \begin{tabular}{|c||c|c|c|c|c|c|}
        \hline
        & $H$ & $b$ & $t_{f}$ & $N$ & $M$ & $\ell_{max}$\\
        \hline 
        \hline 
        Ensemble 1 & $0.5$ & $6.0$ & $2\cdot10^{5}$ & $10^{6}$ & $10^{5}$ & $2\cdot10^{4}$\\
        \hline 
        Ensemble 2 & $0.8$ & $16.0$ & $3\cdot10^{4}$ & $10^{6}$ & $10^{5}$ & $8\cdot10^{4}$\\
        \hline 
        Ensemble 3 & $0.33$ & $3.8$ & $5\cdot10^{7}$ & $10^{6}$ & $10^{5}$ & $2\cdot10^{4}$\\
        \hline 
    \end{tabular}
    \caption{\label{tab1}
        Simulation parameters used to generate the SLE traces. $H$ is the Hurst
        exponent and $b$ is the diffusion coefficient of the fractional
        Brownian motion used as driving function. The curves were
        computed for $N$ times $t_i$ equally spaced in the
        interval $[0, t_f]$. The resulting trace is reparametrized as a function
        of its length and interpolated in $M$ points equally spaced in the interval
        $[0,\ell_{max}]$.
    }
\end{table}

\begin{acknowledgments}
    We thank the Brazilian agencies CNPq, CAPES, and
    FUNCAP, the National Institute of Science and Technology for
    Complex Systems in Brazil, and the European Research 
    Council (ERC) Advanced Grant No. 319968-FlowCCS
    for financial support.
\end{acknowledgments}

\bibliography{paper.bib}

\end{document}